# Machine learning assisted phase and size-controlled synthesis of iron oxides

Juejing Liu[a,b,c,1], Zimeng Zhang[a,d,1], Xiaoxu Li[a,1], Meirong Zong[a], Yining Wang[a], Suyun Wang[a], Ping Chen[a], Zaoyan Wan[a], yatong Zhao[a], Lili Liu[a], Yangang Liang[e], Wei Wang[e], Zheming Wang[a], Shiren Wang[d], Xiaofeng Guo[b], Emily G. Saldanha[f*], Kevin M. Rosso[a*], Xin Zhang[a*]

[a] Physical and Computational Sciences Directorate, Pacific Northwest National Laboratory, Richland, WA 99354, USA

[b] Department of Chemistry, Washington State University, Pullman, WA 99164, USA

[c] Materials Science and Engineering Program, Washington State University, Pullman, WA 99164, USA

[d] Department of Industrial and Systems Engineering, Texas A&M University, College Station, TX 77843, USA

[e] Energy and Environment Directorate, Pacific Northwest National Laboratory, Richland, WA 99354, USA

[f] National Security Directorate, Pacific Northwest National Laboratory, Richland, WA 99354, USA

[1] These three authors contribute equally to this paper.

* Corresponding Author Email: xin.zhang@pnnl.gov (X.Z.), kevin.rosso@pnnl.gov (K.R.), emily.saldanha@pnnl.gov (E.S.)

**Abstract**

    The controllable synthesis of iron oxides particles is a critical issue for materials science, energy storage, biomedical applications, environmental science, and earth science. However, synthesis of iron oxides with desired phase and size are still a time-consuming and trial-and-error process. This study presents solutions for two fundamental challenges in materials synthesis: predicting the outcome of a synthesis from specified reaction parameters and correlating sets of parameters to obtain products with desired outcomes. Four machine learning algorithms, including random forest, logistic regression, support vector machine, and k-nearest neighbor, were trained to predict the phase and particle size of iron oxide based on experimental conditions. Among the models, random forest exhibited the best performance, achieving 96% and 81% accuracy when predicting the phase and size of iron oxides in the test dataset. Premutation feature importance analysis shows that most models (except logistic regression) rely on known features such as precursor concentration, pH, and temperature to predict the phases from synthesis conditions. The robustness of the random forest models was further verified by comparing prediction and experimental results based on 24 randomly generated methods in additive and non-additive systems not included in the datasets. The predictions of product phase and particle size from the models are in good agreement with the experimental results. Additionally, a searching and ranking algorithm was developed to recommend potential synthesis parameters for obtaining iron oxide products with desired phase and particle size from previous studies in the dataset. This study lays the foundation for a closed-loop approach in materials synthesis and preparation, from suggesting potential reaction parameters from the dataset and predicting potential outcomes, through conducting experiments and analysis, to enriching the dataset.

**Introduction**

Over the past few decades, the controlled synthesis of nanomaterials with well-defined size and phase has gained immense attention due to their unique properties and potential applications in catalysis, energy storage, biomedicine, and environmental remediation.[1,2] The size and phase of nanomaterials are essential factors that determines their optical, electronic, magnetic, and catalytic property, consequently, their performance in diverse applications.[3-6] For instance, the size-dependent behavior of nanoparticles, such as the quantum confinement effect in semiconductor nanocrystals, can be harnessed in applications like solar cells, light-emitting diodes, and other optoelectronic devices.[7-9] Smaller nanoparticles have a higher surface-to-volume ratio, leading to a higher number of surface atoms and, therefore, can also enhanced reactivity and catalytic activity.[10-12] On the other hand, phase-controlled synthesis of nanomaterials is equally important, as different phases may exhibit distinct properties and behavior.[1-3, 11] The phase of a material can profoundly influence its catalytic activity and selectivity, as different crystal structures may expose unique active sites and alter the adsorption and desorption kinetics of reactants and products.[3] Furthermore, phase-dependent properties such as electrical conductivity, magnetism, and mechanical strength can significantly impact a material's performance in applications like energy storage, sensing, and drug delivery.[13, 14] Achieving control over both size and phase of nanomaterials is of paramount importance to fully exploit their potential.

Iron oxide nanoparticles, in particular, have attracted considerable interest due to their outstanding magnetic, electrical, and catalytic properties.[15] Among the various iron oxide phases, hematite ($\alpha$-$Fe_2O_3$) has been the subject of extensive research due to its unique characteristics, such as its stability, nontoxicity, photoelectric properties and potential applications in photocatalysis, gas sensing, and antibacterial and nanofluid applications.[16-18] However, achieving precise control over the formation of hematite nanoparticles, as opposed to other iron oxide phases (e.g., maghemite $\gamma$-$Fe_2O_3$, goethite ($\alpha$-FeOOH), akageneite ($\beta$-FeOOH), lepidocrocite ($\gamma$-FeOOH), magnetite ($Fe_3O_4$), and ferrihydrite), and their size remains challenging, owing to the complexities involved in the synthesis process and the inherent sensitivity of these materials to synthesis conditions.[19] In recent years, significant progress has been made in developing various synthesis strategies for iron oxides, including hydrothermal synthesis, sol-gel method, and co-precipitation.[20-22] However, traditionally, to synthesis iron oxide with desired properties, one not only needs to summarize potential synthesis conditions from previous studies but also relies on a trial-and-error process to evaluate the idea, which requires expertise and time. A robust and accurate predictive model for optimizing synthesis protocol of iron oxide to obtain desired phase and particle size is still lacking, which also limits our deeper understanding of the fundamental mechanisms involved in iron oxide nanoparticle formation.

Machine learning (ML) offers a promising avenue to address these challenges by leveraging its capacity for pattern recognition, prediction, and optimization. Such predictive capability is created by incorporating computer algorithms and mathematical models to reveal the underlying relationships between features and labels (in this case, synthesis conditions and morphological parameters).[23, 24] The ML models are particularly helpful when the relationships between variables and the outcomes are complicated or unknown.[25, 26] With the expansion of experimental data sources and technological advances for high-throughput experimental setups, the ML models can take advantage of the large amount of data available. And these models have proven to be effective in establishing the synthesis-structure-property relationships of desirable materials and exploring the new structures.[27-29] Various ML models have been proposed to be used in analysis of different nanomaterials and their specific properties, such as logistic regression (LR),[30] random forest (RF),[31] k-nearest neighbor (KNN),[32] Gaussian processes,[33] support vector machines,[34] deep neural network (DNN),[35] and Bayesian optimization.[36] For example, Sun et al. combined various ML analytical models and physics-based simulation to predict the electron transfer to achieve efficient and accurate high-throughput production for silver nanoparticles, among studies of other metal particples.[37] Wang et al. developed an unsupervised ML model for transmission electron microscopy (TEM) image analysis of gold nanoparticles.[38] The morphological information was extracted from the TEM images, followed by a rapid classification of particle without human intervention. The relationships between the synthesis conditions and the morphology of the nanoparticles were determined. Lee et al. also used ML to quantitively analyze the morphology of gold nanoparticles via TEM images and obtained a high precision of 99.75%.[39] Pellegrino et al. studied the performance of ML model for predicting the morphology of $TiO_2$ nanoparticles obtained with different synthesis parameters.[40] The previous studies show the potential of using ML to reveal the underlining relationship between synthesis conditions and the properties of iron oxide nanomaterials.

Another common encounter question in materials synthesis and preparation is how to choose the right experimental conditions. It is well known that the selection of appropriate synthesis parameters is the key to obtaining materials with desired properties. Traditionally, in many studies, the synthesis parameters have to be manually extracted and summarized to obtain optimal sets of synthesis parameters. Such a process is not only time-consuming, but also requires expertise and strong intuition. Although the recent development of advanced ML-based data mining has efficiently gathered synthesis parameters from thousands of studies,[41] the significant amount of data makes manual verification virtually impossible. Therefore, it is of importance to develop an objective method to evaluate, rank, and recommend the conditions with relatively higher possibility to synthesize desired products. To our knowledge, only a few studies have discussed the recommendation of synthesis from existing data sets.[42, 43]

In this study, based on an iron oxide synthesis dataset we mined from previous studies and unpublished data in our lab, we demonstrate the potential solutions for the two key questions above: Understanding potential outcomes from sets of synthesis parameters, and obtaining the viable synthesis parameters from specified products. For the first question, we trained and tested different ML models for predicting the synthesis results is or is not hematite phase and the particle size if resulting is hematite according to the synthesis condition, including logistic regression from generalized linear models, decision tree based random forest, voting based k nearest neighbor, and instance based support vector machine. The premutation feature importance analysis was also preformed to show which are most important features that models rely on to predict the particle size and phase from synthesis conditions. Correlation analysis was preformed to study the relations between important features. The robust of random forest model were further verified by comparing prediction and experimental results, based on 24 randomly generated methods in additive-added and additive-free system which were not included in the datasets. The predictions of product phase and particle size from models are good agreed with the experimental results. For the second question, we designed a searching and ranking algorithm to find potential parameters to synthesize iron oxides with desired phase and particle size. The combination of these two solutions can potentially become a "close loop" method encompassing selection of parameter from dataset, prediction of plausible outcome, conduct experiment and evaluation, and enrich dataset.

**Experiments and Methods**

**Chemicals and Materials.** Ferric chloride anhydrous, sodium dodecyl sulphate (SDS), and sodium citrate, and sodium hydroxide were purchased from Sigma-Aldrich Chemical Reagent Co., Ltd. All chemicals are analytical purity and can be used directly without any further treatment. Deionized (DI) water used in this work was prepared using a Barnstead water purification system.

**Synthesis of Iron Oxides.** In a typical procedure, anhydrous ferric chloride and additives were dissolved in 15 ml of DI water under magnetic stirring at room temperature. The pH of the solution was then adjusted to the desired value by slowly adding 5 M NaOH and monitoring with a Thermo Scientific Orion Star A221 pH meter. The resulting solution was transferred into a 20 mL Teflon liner stainless steel autoclave and kept at the specific temperature and time. The products were washed three times by DI water and the final products were collected by centrifugation at 8000 rpm.

**Characterization.** Powder XRD of all as-prepared samples was performed on a Philips X'pert Multi-Purpose Diffractometer (MPD) (PANAlytical) equipped with Cu Kα radiation operating at 50 kV and 40 mA. Scanning Electron Microscopy (SEM) imaging was conducted on the FEI Helios NanoLab 600i dual-beam focused ion beam precision manufacturing instrument operating at 5 kV and 86 μA. To improve the electronic conductivity of the samples before SEM imaging, a thin carbon layer (about 12 nm) was deposited on the particle surface by using a carbon coater (208C; Ted Pella, Inc.).

**Data Set Characterization and Collection.** Data on the synthesis methods, phase and particle size of iron oxide nanomaterials have been collected from previously reported studies as well as experimental data collected in the laboratory at Pacific Northwest National Laboratory, USA. The dataset includes contains 700 pieces of data corresponding to iron oxides synthesis in different sets of conditions, including precursors, additives, solvents, concentrations of each ingredient and temperature. The corresponding phases of the iron oxides such as hematite ($\alpha$-$Fe_2O_3$), maghemite ($\gamma$-$Fe_2O_3$), magnetite ($Fe_3O_4$). Goethite ($\alpha$-FeOOH), akageneite ($\beta$-FeOOH), lepidocrocite ($\gamma$-FeOOH), ferrihydrite and the size of the nanoparticles were also included in the dataset. The dataset was consisted of about 500 sets of data from previous publications reporting the iron oxides synthesis from 2010 to 2020 as well as the 200 sets of laboratory experimental data.

**Software Libraries.** All software libraries used in this study are listed below: Pandas,[44] NumPy,[45] scikit-learn,[46] openpyxl,[47] and Jupyter Notebook.[48] Pandas, NumPy, and openpyxl were employed for the purpose of importing, cleansing, and preprocessing data. The scikit-learn library and Jupyter Notebook were employed to train models based on different algorithms in an interactive Python environment. Data visualization and cluster analysis were conducted using Seaborn and Matplotlib.[49, 50]

**Feature Analytics and Selection**. Eleven features were selected based on our experimental experience as having potential to significantly impact the synthesis of iron oxides. These features include precursor species and concentration, surfactant species and concentration, hydrogen ion concentration, temperature, reaction time, solvent, and solvent volume. Among the features selected for this study, precursor and surfactant species, as well as solvent, were categorical variables. To prepare the data for training the machine learning models, one-hot encoding was applied to each categorical variable. This process created additional feature columns from each categorical variable, with each unique value represented as a binary feature. For instance, the precursor variable contained several unique values, including $FeCl_3$, $Fe(NO_3)_3$, $FeSO_4$, $FeC_2O_4$, $K_3[Fe(CN)_6]$, etc. Consequently, 10 additional feature columns were generated to replace the original precursor feature column. A value of 1 in a particular precursor column indicated the use of that precursor in an experiment, while 0 indicated the use of a different precursor. Normalization was performed on the training set using the StandardScaler class from scikit-learn. The resulting standard scaler was then applied to both the training and test sets. The purpose of this normalization was to minimize any potential bias towards the test set, which should be considered as unknown during the training phase. By normalizing both the training and test sets in the same way, we aimed to ensure that the machine learning models could make accurate predictions on previously unseen data.

**Stratification, Sampling, and Training**. In our dataset, hematite was the primary synthesis products. However, to address the issue of dataset imbalance, we used stratified sampling to ensure a relatively equal distribution of hematite and non-hematite samples in both the training and testing subsets. Specifically, we allocated 80% of the data to the training subset and 20% to the testing subset. This approach aimed to improve the robustness of the machine learning models by training them on a representative sample of the data and testing their generalization ability on previously unseen data.

We trained models based on four ML algorithms, namely k-nearest neighbor (KNN), logistic regression (LR), support vector machine (SVM), and random forest (RF).[30-32, 34] To achieve high accuracy in our analysis, k-fold cross-validation was then used in this study. We split the training dataset into k (in this study, k = 5) equally sized subgroups. n each validation subgroup, we used the confusion matrix to measure accuracy, while the remaining four subgroups were used for training the machine learning models. The k-fold cross-validation process was repeated five times for each algorithm, with each subgroup used only once for validation. The mean value across the k-folds was calculated, and the cross-validation process was iterated for various combinations of hyperparameters. During the training of the machine learning models for all four algorithms, we performed grid search cross-validation to evaluate multiple models with different training parameters (see Table S1). We then selected the best model for each algorithm and compared their performance. To evaluate the performance of the models, we used the testing dataset, which was not used during the training process. We measured the prediction performance using accuracy, which is defined as the ratio of correctly predicted data to the total testing data.

To obtain lists of feature importance for each algorithm after training, we utilized the permutation method available in scikit-learn. This method involves shuffling each feature per epoch and evaluating the resulting impact on model accuracy. Features that have a significant effect on model accuracy when shuffled are considered to be of high importance, while those that do not significantly affect model accuracy are considered to be of lower importance. We then performed a correlation analysis on the five most important features (temperature, precursor concentration, pH, time, and solution volume) using Pearson's correlation coefficients. Additionally, a hierarchically clustered heatmap was generated using seaborn to observe the relationship between these five conditions and the phase of the iron oxide products. We converted the prediction of whether the experimental conditions led to the formation of hematite or not into a binary classification. Two methods were employed to predict the size of nanoparticles, with regression based solely on the RF algorithm and classification based on all four algorithms. In the classification method, we sorted the nanoparticle sizes into three categories: nano (less than 100 nm), sub-micron (between 100 nm and 1000 nm), and micron (greater than 1000 nm). We excluded LR from the classification method since it is only capable of binary classification.

**Design of searching and ranking recommendation algorithm.** The recommendation algorithm took the desired phase and particle size of iron oxide as input. The ranking algorithm allowed us to quantify the degree of deviation of each parameter from its corresponding average value, using the standard deviation as a measure of variability, to recommend the most suitable conditions for achieving the desired phase and particle size of iron oxide. It then searched the entire dataset and selected all possible sets of conditions that met the desired criteria. We calculated the average (avg.) and standard deviation (std.) of the major features, including time, temperature, pH, precursor concentration, and volume, from the selected sets of parameters. To evaluate how a specific parameter of a specific set differs from the averaged value, we used a ranking algorithm as shown in Equation 1:

$$s_{feature} = \frac{\sigma_{feature}}{|V_{feature} - avg_{feature}|} \quad (Eq.\ 1)$$

, where $s_{feature}$ is the score of the specific parameter in the specific set, $V_{feature}$ is the raw value of the feature, $avg_{feature}$ and $\sigma_{feature}$ is the average and standard derivation of the feature. The scores from every parameter in the set were then added up and form the final score of the set of features:

$$S_{set} = \sum_i s_{feature,i} \quad (Eq.\ 2)$$

A high score ($S_{set}$) indicates that the parameters in this set are generally closer to the averaged value and more previous studies used similar parameters to synthesis the desired products. In contrast, a low score suggests that the parameters are farther from the averaged values and less studies used similar parameter to synthesis the desired products.

**Results and Discussion**

**Machine learning models.** We collected a total of 792 sets of data corresponding to 792 different iron oxide synthesis experiments (see Figure 1a). Each set of data contains a list of experiment conditions as features, including type and concentration of reactants, temperature, pH value, reaction time, etc. Each set of experimental conditions corresponds to a series of resultant parameters as labels, such as the phase and size of iron oxide particles. In general, training ML models requires significant amount of data.[51-54] The small size of the dataset prevents us from using more sophisticated algorithms, such as neural networks. Therefore, we chose four algorithms that can handle our small dataset. We trained various machine models, including k-nearest neighbor (KNN),[32] logistic regression (LR),[30] support vector machine (SVM),[34] and random forest (RF),[31] to determine the formation of hematite phase and the size of the nanoparticles (Figure 1b). KNN is a classical instance-based classification algorithm. During the training process, KNN classifies experimental conditions (features) in the training dataset into several different outcomes (labels). When an unseen feature is fed to the trained model, the model evaluates the similarity of the unseen feature to all known features. If the model detects that such an unseen feature is similar to the known features with a particular label, then the model classifies the unseen feature into that label. LR is one of the generalized linear models and is an exclusive binary classification algorithm. During training, the LR algorithm builds a function to calculate the probability of features belonging to one bin or another. When the LR model encounters an unseen feature, it uses the training function to calculate the probability and then categorize the feature. The SVM model is a kernel-based ML algorithm. It transforms the features into a higher dimension by the kernel algorithm (in our case, radial kernel) and then generates rules to classify the features. SVM models contain hundreds of decision trees and combine the result of each decision tree to determine the final prediction. We first examined the performance of the models from the four algorithms with a binary classification question: Whether the experimental conditions lead to hematite formation or not.

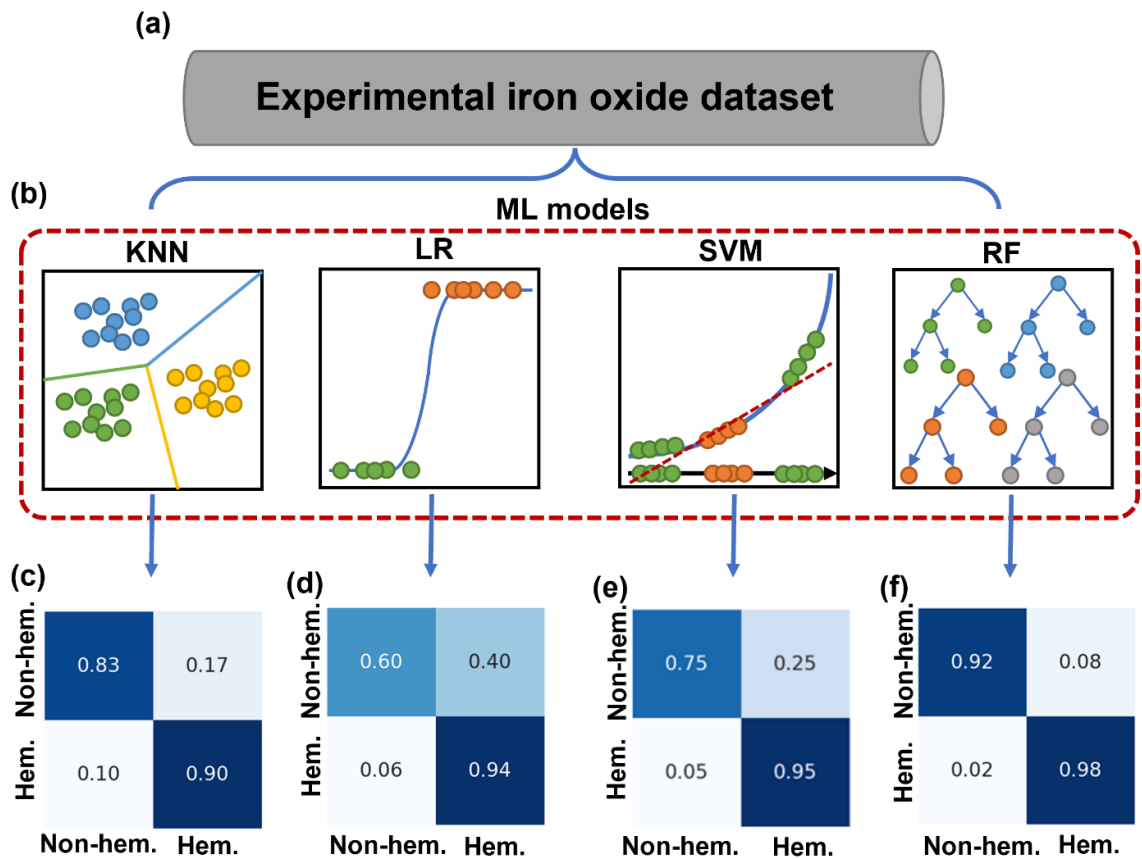

**Figure 1.** Prediction of formation of hematite based on experimental conditions by different ML methods. (a) A dataset of iron oxide synthesis conditions obtained from previous studies and our experimental results. (b) ML algorithms used in this study, including k-nearest neighborhood (KNN), logistic regression (LR), support vector machine (SVM), and random forest (RF). (c) to (f) Utilizing models from KNN (c), LR (d), SVM (e), RF (f) to predict the formation of hematite.

For this binary classification question, the RF-based model outperformed all three other models. The KNN based model predicts the formation of a non-hematite compound with 83% accuracy, while the accuracy for predicting the formation of a hematite compound is 90%. The prediction accuracies of the LR model are more divergent than those of the KNN model. The accuracy for detecting the formation of non-hematite is only 60%. This low accuracy is close to random guessing (50%). In contrast, the model achieves 94% accuracy in predicting the formation of hematite compounds. The SVM model achieves 75% accuracy in detecting the formation of non-hematite and 95% accuracy in predicting the formation of hematite. The RF model has the highest accuracy of the four algorithms. The accuracy of predicting non-hematite formation is 92%, while the accuracy of predicting hematite formation is 98%. The RF model also achieves the highest overall accuracy of all four models (96%, see Table S2).

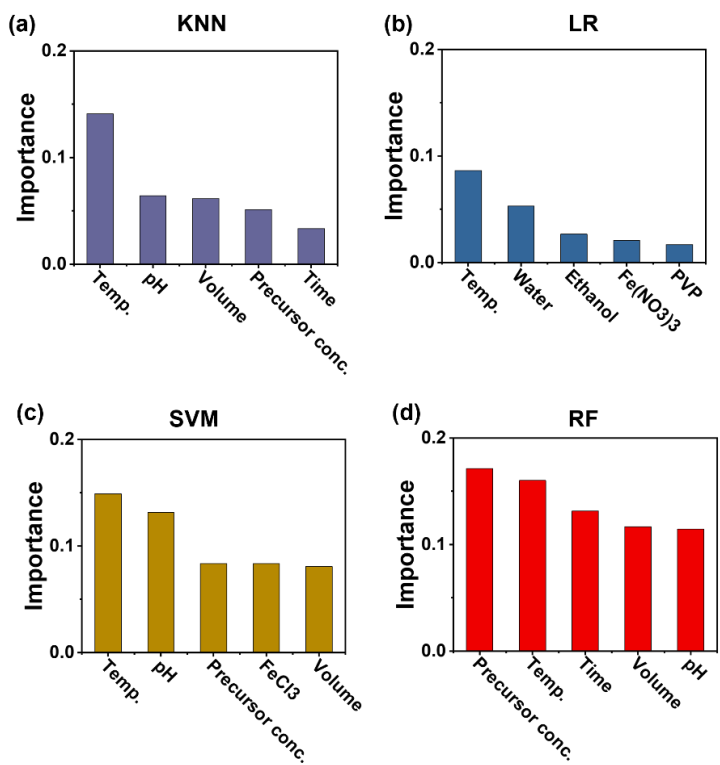

**Figure 2.** Premutation feature importance of k-nearest neighborhood (a), logistic regression (b), support vector machine (c), and random forest (d).

The feature importance results show that most models rely heavily on the features including temperature, pH, precursor concentration and time, to predict the formation of hematite (Figure 2). The feature importance lists for the four models were obtained using a permutation method. This method shuffles the values in one feature while leaving other features intact, and then observes the effect of the shuffled feature on the accuracy of the models. If the shuffled future has a significant impact on accuracy, then that feature would be considered to be of high importance, and vice versa. In the KNN model (see Figure 2a), temperature is the most important feature, followed by pH, volume, precursor concentration, and time. It is surprising that the model considers volume as an important feature, especially when precursor concentration is simultaneously present. A plausible reason is that the majority of our data set is hydrothermal synthesis using autoclaves. The ratio of solution to autoclave volume determines the pressure inside the autoclave.[55] As it is well known that pressure is one of the most important features in hydrothermal reactions,[56] the feature of volume is therefore chose by the algorithms to determine the formation of hematite.

The LR model is the only one that shows a different set of feature importance ranks than those in the other models. Although temperature is still the most important feature, the ranking of the other features is less intuitive. After temperature, this model relies on whether water and ethanol are used as solvents, or whether Fe(NO$_3$)$_3$ and polyvinylpyrrolidone (PVP) are used during the synthesis. In addition, the LR model relies less on these features to predict hematite formation. The value of the importance of temperature is about two thirds that of the KNN models (0.09 vs. 0.14). Hence, we believe that the LR model does not recognize the pattern of iron-oxide synthesis, especially for the non-hematite part (see Figure 1d and Table S2).

The SVM model relies on similar features as the KNN model to predict the outcome of the iron oxide synthesis (see Figure 2c). However, the feature of the presence of FeCl$_3$ precursor is more important than the reaction time. The SVM model also relies more on the first five features to predict the phase of the iron oxide than KNN, as the values of importance of the first five features are higher than SVM model. As a result, the SVN model shows a higher accuracy in terms of binary phase prediction (Table S2).

The RF model not only relies on the commonly known important features (precursor concentration, temperature, time, volume, and pH) but also seems to treat them equally. The highest and lowest values of importance are 0.17 for precursor concentration and 0.11 for pH, respectively. Mixing any of these five features has a similar effect on accuracy of the RF model. Therefore, the accuracy of the RF model is the highest of all four models.

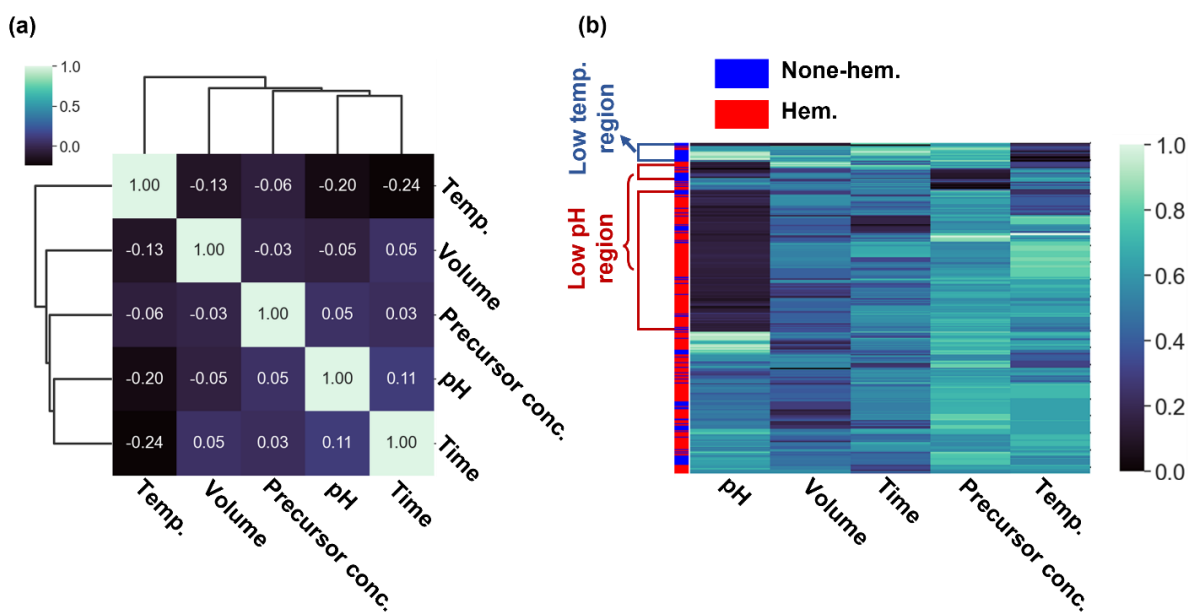

**Figure 3.** Correlation analysis of important features, including temperature, volume, precursor concentration, pH value, and time. The dendrogram with Pearson correlation coefficient (a) and cluster map (b) were generated based on the whole iron oxide synthesis dataset. Data was standardized between 0 and 1 to amplify differences.

We further preformed the correlation analysis using the Pearson correlation coefficient for the five main features determined by the premutation method (temperature, volume, precursor concentration, pH, and time). The result shows that most of the features are independent from each other (see Figure 3a), although a dendrogram was still obtained. Most of the correlation coefficients are smaller than 0.1 indicating that negligible correlation. This is understandable as iron oxides can be synthesized in many different combinations of conditions.[57] The first exception is temperature, which has a relatively strong negative correlation with the other features. One possible explanation for the correlation between temperature and solution volume could be related to the synthesis method used in the majority of the experiments in our dataset. Specifically, most of the syntheses used the hydrothermal method with autoclaves. In this method, high temperatures can lead to a significant increase in pressure inside the autoclave. If the volume of the solution is also high, this can exacerbate the pressure increase and may even cause leaking. As a result, the experimental conditions used in these experiments may have been limited by the maximum allowable temperature and volume for the given autoclave setup. This could explain why there is a negative correlation between temperature and solution volume in our dataset. However, further experiments are needed to confirm this hypothesis.[55] Additionally, the negative correlation between temperature and time can likely be explained by the reaction kinetics of iron oxide synthesis. Increasing the temperature during the synthesis process is generally favorable for the formation of iron oxide crystals, which can shorten the reaction time required to achieve a desired outcome.

The hierarchically clustered heatmap (as shown in Figure 3b) provided insight into the relationship between pH and temperature and the formation of hematite or other iron oxide phases, which is consistent with the principle of classical nucleation theory.[19] Our results show that, under low pH (below 4.0), most of the synthesized products were hematite (indicated in red). Conversely, at higher pH levels (above 4.0), the proportion of non-hematite products (indicated in blue) increased. The reason for this preference for acid solutions in hematite synthesis is since hematite is thermodynamically stable under low pH conditions compared with other metastable iron oxides phase. In addition, acidic conditions can lead to the dissolution of iron-containing precursors and promote the nucleation and growth of hematite crystals. However, the oversaturation state of solution in terms of hematite decreases with decreasing of the solution pH, therefore a higher temperature is desired to overcome the thermodynamical barrier of nucleation to form the stable hematite phase. Therefore, while low pH conditions are beneficial for hematite synthesis, a higher temperature is needed to promote the nucleation and growth of hematite crystals.[19, 58] In the higher pH range, the solution is oversaturated in terms of all ferric oxides. In such case, the surface energy will perform a higher impact on the phase selection. Because the metastable phase has lower surface energies than thermodynamical stable phase hematite, the barrier to form a critical nucleus of metastable phase is lower.

The metastable phase will form first and simultaneously depletes the concentration of precursor, which hinders the formation of hematite at relatively high pH conditions.

Particle size is an important factor that can significantly impact the catalytic performance of iron oxide products.[22, 59-61] Therefore, it is crucial to know the particle size of the particles before conducting the experiment. Therefore, predicting the particle size of synthesized iron oxides based on the reaction conditions is of great interest. In this study, we initially attempted to use the random forest (RF) regression algorithm to train models for predicting the exact particle size based on the experimental conditions. However, as shown in Figure S1, the RF regression model was unable to accurately predict the particle size of iron oxide particles from the test dataset. We hypothesize that the small size of our dataset may have contributed to the low accuracy of the RF regression model, as there may not have been sufficient information to train the model effectively. To address this issue, we converted the prediction of particle size from a regression question into a classification question. By sorting the particle size into three categories (nano, sub-micron, and micron), we were able to use all four ML algorithms to train models for predicting particle size based on the experimental conditions.

After converting the particle size prediction task from regression to classification, we trained models based on three ML algorithms: support vector machine (SVM), k-nearest neighbor (KNN), and random forest (RF). Logistic regression was excluded from the analysis due to its binary classification nature. As shown in Figure 4a, we divided the particle size into three categories: nano (smaller than 100 nm), sub-micron (between 100 nm and 1000 nm), and micron (larger than 1000 nm).

During the training process, the SVM algorithm failed to converge and produced no model (Figure 4b). Although the KNN algorithm converged during training, the resulting model had a relatively low overall accuracy (62%, Figure 4c), with accuracies of 61% (nano), 63% (sub-micron), and 67% (micron) for predicting particle sizes from conditions in the test dataset. In contrast, the RF-based model demonstrated the highest accuracy among the three algorithms, with an overall accuracy of 81% (Figure 4c). Specifically, the RF-based model achieved accuracies of 80%, 83%, and 75% for predicting nano, sub-micron, and micron particles, respectively.

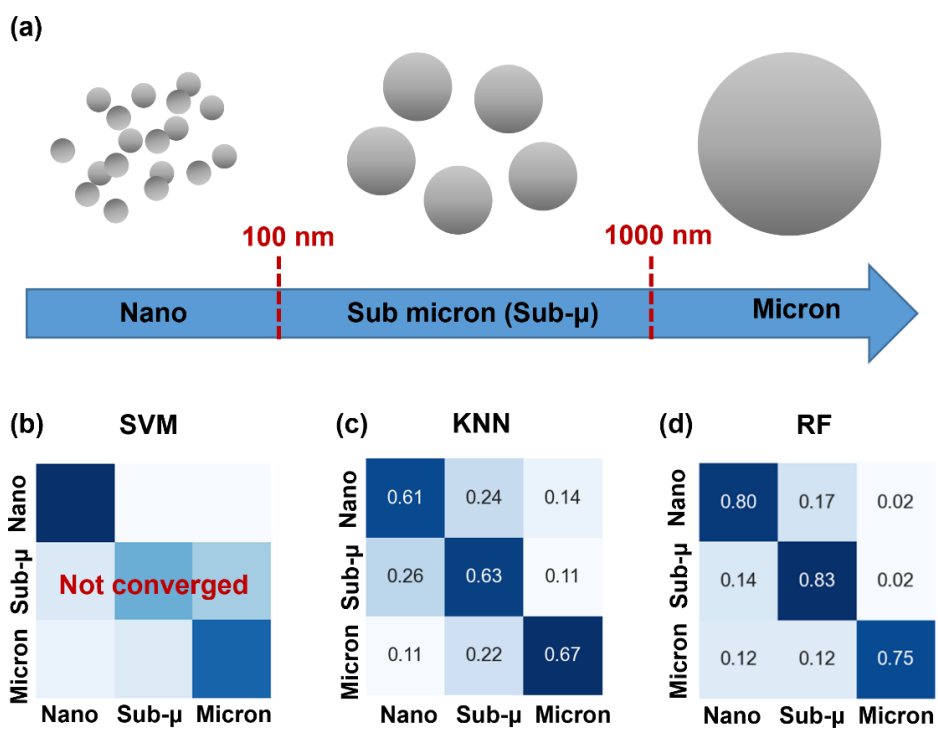

**Figure 4.** Prediction of particle size of products based on conditions of reactions. (a) Dividing product particles into three categories based on size. (b) to (d) Prediction results based on models from SVM (b), KNN (c), and RF (d). Training SVM models were unsuccessful since fitting was not converged.

**Experimental validation of ML models**. To further evaluate the performance of the random forest-based binary classification model and particle size prediction model, we conducted experiments to synthesize iron oxides with and without adding surfactant. We prepared 18 samples without adding additive (see Table 1, Figure S2 and S3). The binary phase classification model correctly identified the formation of hematite from 15 sets of conditions, resulting in an accuracy of 83%. However, the model incorrectly classified three sets of conditions, which were the combination of high pH with high temperature (Exp. 3, pH at 12.6 and temperature at 180 °C) or low pH with low temperature (Exp. 4 and 5, pH at 1.1 and 2.0, temperature at 80 °C). A similar trend was observed in the cluster map (Figure 3b), where low pH or high reaction temperature was found to cause the formation of hematite. Therefore, this binary phase classification model relies on one of these two factors, high temperature or low pH, to determine the formation of phases in some cases.

Based on SEM observations, it appears that the RF model for predicting particle size based on experimental conditions is less accurate than the binary phase classification model. In particular, the RF model correctly predicted the size range of particles in five out of eight samples (Exp. 1, 2, 3, 8, and 15). However, for samples 7 and 13, the model's predictions were mixed, as SEM images showed that sample 7 contained both nano and sub-micron particles, while sample 13 was made up of sub-micron and micron particles. The model also incorrectly predicted the particle size of sample 14 as sub-micron, when SEM observations showed it contained nano rods.

Table 1. Utilizing random forest-based models to predict phase and size of iron oxide products on random generated experiments. Volume of solvent and reaction time were fixed at 15 ml and 16 hours, respectively. XRD was used to identify phases (Figure S2 and S3), and SEM was utilized to identify size of particles (Figure S4)

| No. | FeCl$_3$ conc. (mM) | pH | Temp. (°C) | Is hem. Predict | Predict size | Exp. phase* | Exp. size |
|---|---|---|---|---|---|---|---|
| Exp 1 | 500 | 1.1 | 180 | True | Micron | Hem | Micron |
| Exp 2 | 428 | 2.0 | 180 | True | Micron | Hem | Micron |
| Exp 3 | 372 | 12.6 | 180 | True | Nano | Gt | Nano |
| Exp 4 | 500 | 1.1 | 80 | True | Micron | Gt | |
| Exp 5 | 428 | 2.0 | 80 | True | Micron | Gt | |
| Exp 6 | 372 | 12.6 | 80 | False | Nano | Gt | |
| Exp 7 | 100 | 1.7 | 180 | True | Sub-μ | Hem | Sub-μ & Micron |
| Exp 8 | 98 | 2.1 | 180 | True | Sub-μ | Hem | Sub-μ |
| Exp 9 | 90 | 12.8 | 180 | False | Sub-μ | Gt | |
| Exp 10 | 100 | 1.7 | 80 | False | Sub-μ | Aka | |
| Exp 11 | 98 | 2.1 | 80 | False | Sub-μ | Aka | |
| Exp 12 | 90 | 12.8 | 80 | False | Nano | Gt | |
| Exp 13 | 10 | 2.4 | 180 | True | Sub-μ | Hem | Nano & Sub-μ |
| Exp 14 | 10 | 11.9 | 180 | False | Sub-μ | Gt | Nano |
| Exp 15 | 10 | 3.7 | 180 | True | Sub-μ | Hem | Sub-μ |
| Exp 16 | 10 | 2.4 | 80 | False | Sub-μ | Aka | |
| Exp 17 | 10 | 11.9 | 80 | False | Nano | Gt | |

In the additive group, the RF model accurately predicted the phase and size of iron oxide particles synthesized with sodium dodecyl sulfate (SDS), while the predictions for all experiments involving sodium citrate (SC) were inaccurate (refer to Table 2, Figure S5, and S6). A closer examination of the experimental conditions for iron oxide synthesis with SDS (Exp. 19 to 21) revealed that the additive may alter the morphology of the iron oxide particles, but not the phase type. The synthesis conditions of Exp. 19 and 21 were in acidic or weak basic environments (pH of 2.0 and 8.0, respectively) with relatively high reaction temperature (180 °C). The combination of these two conditions is known to cause the formation of hematite. In Exp. 20, the reaction temperature was only 110 °C, favoring the formation of non-hematite products, such as akageneite in this case. On the other hand, the introduction of SC significantly altered the phase of the iron oxide products, even when the experimental conditions clearly favored the formation of hematite. For instance, in Exp. 22, we observed the formation of Magnetite ($Fe_3O_4$) at high reaction temperature and low pH value. Moreover, a previously unseen iron oxide phase, ferrihydrite, was formed in the low-temperature reaction involving SC. The reason SC significantly altered the formation of phases is that this additive can cause the reduction of $Fe^{3+}$ ions into $Fe^{2+}$. In contrast, SDS additive does not react with $Fe^{3+}$ ions. Overall, the random forest model exhibited good predicting accuracy when the additive was not reactive with Fe ions.

**Table 2.** Utilizing random forest-based models to predict phase and size of iron oxide products on random generated experiments with surfactant. Volume was fixed at 15 ml. XRD was used to identify phases (see Figure S5), and SEM was utilized to identify size of particles (see Figure S6)

| No. | $FeCl_3$ conc. (mM) | Time (h) | pH | Additive* | Additive conc. (mM) | Temp. (°C) | Is hem. Predict | Predict size | Exp. phase** | Exp. size |
|---|---|---|---|---|---|---|---|---|---|---|
| Exp 19 | 831 | 7 | 2.0 | SDS | 131 | 180 | True | Nano | Hem | Nano |
| Exp 20 | 971 | 5 | 8.0 | SDS | 181 | 110 | False | Nano | Aka | Nano |
| Exp 21 | 501 | 29 | 8.0 | SDS | 121 | 180 | True | Sub-µ | Hem | Sub-µ |
| Exp 22 | 461 | 37 | 2.0 | SC | 141 | 180 | True | Sub-µ | Mag | Nano |
| Exp 23 | 671 | 41 | 2.0 | SC | 191 | 95 | True | Micron | 2L-Fh | Nano |
| Exp 24 | 471 | 34 | 10.0 | SC | 61 | 110 | False | Nano | 2L-Fh | Nano |

*SDS: Sodium dodecyl sulfate, SC: Sodium citrate.
** Hem: Hematite, Gt: Goethite, Aka: Akaganeite, Mag: Magnetite, 2L-Fh: 2-line Ferrihydrite.

**Optimizing synthesis algorithm based on dataset.** We developed a search and ranking algorithm to generate recommendations for synthesizing iron oxide particles with specific desired properties, such as phase and particle size. The algorithm is based on the idea that the values in a feature that are frequently used to synthesize the desired product are more likely to lead to success than those rarely used. To implement this idea, we retrieved all possible sets of reaction conditions from the iron oxide dataset that could produce the desired product and calculated the average and standard deviation of the features from the resulting sub-dataset. We then calculated a ranking score for each set of parameters using a detailed algorithm outlined in the methods section. The algorithm considers the distance between each feature value and its corresponding average value in the sub-dataset. Sets of parameters with higher ranking scores are more likely to have been used in previous studies to synthesize the desired product. Conversely, sets of parameters with lower ranking scores are less likely to have been used. The output of the algorithm is a list of recommended sets of parameters for synthesizing the desired product. This algorithm can significantly reduce the time and effort required to search for specific iron oxide particles. An example of the algorithm's output for synthesizing hematite nanoparticles with diameters between 25 nm and 75 nm is shown in Table S2.

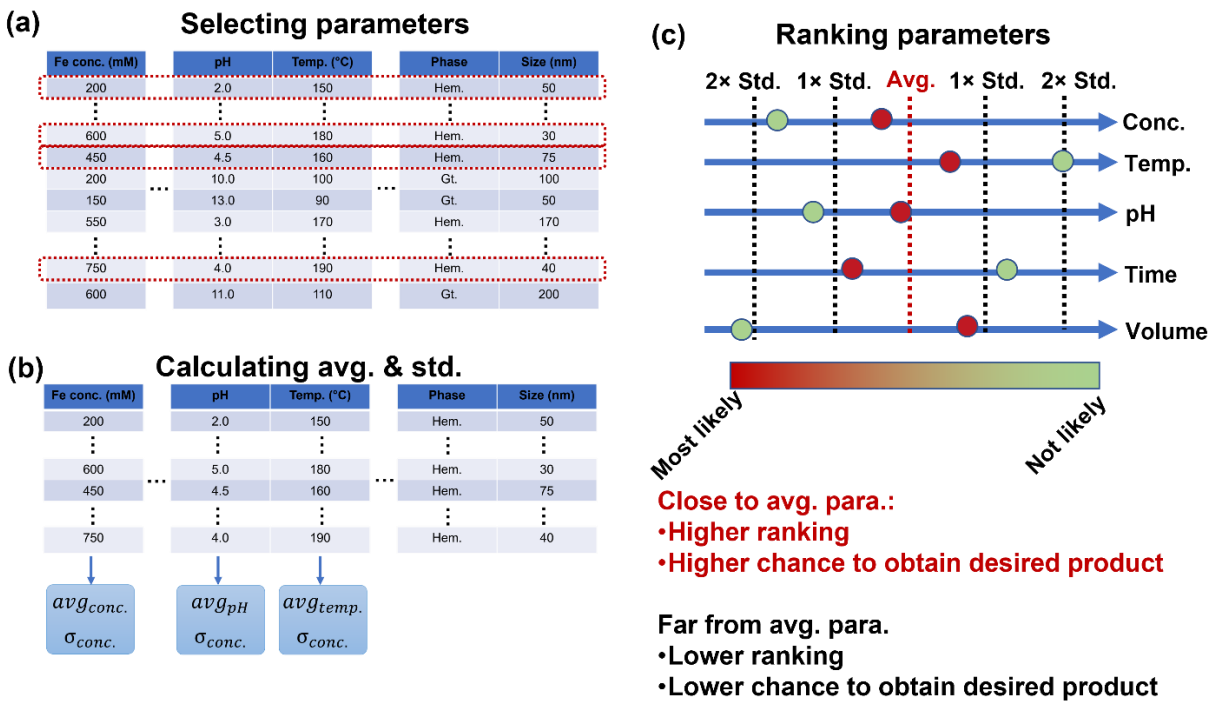

Figure 5. Retrieving desired synthesis parameters to obtain iron oxide particles with specified phase and particle size. (a) Selecting all possible sets of parameters from dataset (e.g., hematite with 25 nm to 75 nm diameter). (b) Calculating average and standard derivation from every category of parameter. (c) Illustration of ranking parameters in every set and summarize result. The red dots show the set of parameters more likely to synthesize desired product, vice versa for the green dots.

## Conclusion

This study presents solutions for two fundamental challenges in materials synthesis: predicting the outcome of a synthesis from specified reaction parameters and correlating sets of parameters to obtain products with desired outcomes. To predict experimental outcomes, four machine learning algorithms, including random forest, logistic regression, support vector machine, and k-nearest neighbor, were trained to predict the phase and particle size of iron oxide based on experimental conditions. Among the models, random forest exhibited the best performance, achieving 96% and 81% accuracy when predicting the phase and size of iron oxides in the test dataset. The permutation feature importance analysis revealed that most models relied on intuitive features such as precursor concentration, pH, temperature, and reaction time to predict the formation of phases based on experimental conditions. The random forest-based models were evaluated for iron oxide synthesis in both additive-free and additive systems, and overall showed good accuracy. Additionally, a searching and ranking algorithm was developed to recommend potential synthesis parameters for obtaining iron oxide products with desired phase and particle size from previous studies in the dataset. This study lays the foundation for a closed-loop approach in materials synthesis and preparation, from suggesting potential reaction parameters from the dataset and predicting potential outcomes, through conducting experiments and analysis, to enriching the dataset.

**Data and model availability**

The models, code and dataset used in this study will be available by requests.

**Acknowledgments**

This material is based upon work supported by the U.S. Department of Energy, Office of Science, Office of Basic Energy Sciences, Chemical Sciences, Geosciences, and Biosciences Division through its Geosciences program at Pacific Northwest National Laboratory (PNNL) (FWP 56674). A portion of the work was performed using the Environmental and Molecular Sciences Laboratory (EMSL), a national scientific user facility at PNNL sponsored by the DOE's Office of Biological and Environmental Research. PNNL is a multi-program national laboratory operated by Battelle Memorial Institute under contract no. DE-AC05-76RL01830 for the DOE. E.S., Y.L., W.W. and X.Z. also acknowledge supported by Energy Storage Materials Initiative (ESMI), which is a Laboratory Directed Research and Development Project at Pacific Northwest National Laboratory (PNNL). S.W. acknowledge the support of this work by the National Science Foundation (NSF), Division of Civil, Mechanical, & Manufact Innovation, under award No. 1934120. X.G. acknowledge the support of this work by the National Science Foundation (NSF), Division of Earth Sciences, under award No. 2149848.

**Ethics declarations**

The authors declare no competing financial interests